\documentclass[runningheads]{llncs}
\usepackage[T1]{fontenc}
\usepackage{adjustbox}
\usepackage{siunitx,booktabs}
\usepackage{xcolor,colortbl}
\usepackage{diagbox}
\usepackage{multirow}
\usepackage{amsmath}
\usepackage{amssymb}
\usepackage{hyperref}
\usepackage{bbding}

\usepackage{graphicx}
\usepackage{color}

\newcommand{\repeatthanks}{\textsuperscript{\thefootnote}}
\begin{document}
\title{Shape-Aware Masking for Inpainting in Medical Imaging}

\iffalse
\author{Yousef Yeganeh\thanks{The first two authors contributed equally to this work.} \\
	%Department of Informatics\\
	Technical University of Munich\\
	Munich, Germany \\
	\texttt{y.yeganeh@tum.de} \\
	%% examples of more authors
	\And
	Azade Farshad\footnotemark[1] \\
	%Department of Informatics\\
	Technical University of Munich\\
	Munich, Germany \\
	\texttt{azade.farshad@tum.de} %\\
	\And
	Nassir Navab \\
	%Department of Informatics\\
	Technical University of Munich\\
	Munich, Germany \\
	\texttt{nassir.navab@tum.de} \\
}
\fi

\author{Yousef Yeganeh\thanks{Equal Contribution}\inst{1} \and Azade Farshad
\repeatthanks \inst{1} \and
Nassir Navab\inst{1,2}
}
\authorrunning{Y. Yeganeh et al.}
\institute{Technical University of Munich, Munich, Germany \and
Johns Hopkins University, Baltimore, USA
}
\maketitle              % typeset the header of the contribution
\begin{abstract}
Inpainting has recently been proposed as a successful deep learning technique for unsupervised medical image model discovery. The masks used for inpainting are generally independent of the dataset and are not tailored to perform on different given classes of anatomy. In this work, we introduce a method for generating shape-aware masks for inpainting, which aims at learning the statistical shape prior. We hypothesize that although the variation of masks improves the generalizability of inpainting models, the shape of the masks should follow the topology of the organs of interest. Hence, we propose an unsupervised guided masking approach based on an off-the-shelf inpainting model and a superpixel over-segmentation algorithm to generate a wide range of shape-dependent masks. Experimental results on abdominal MR image reconstruction show the superiority of our proposed masking method over standard methods using square-shaped or dataset of irregular shape masks.

\end{abstract}
\section{Introduction}
Recent advances in deep learning and especially generative models have made it possible to take advantage of these methods for image generation, completion or manipulation tasks. One of the commonly used image manipulation methods is image inpainting \cite{bertalmio2000image,dhamo2020semantic}, which aims to reconstruct images that are partially masked. The masks used for training an inpainting model can considerably affect its performance. In traditional computer vision tasks, convolutional networks \cite{suvorov2022resolution,yeganeh2020inverse} yield state-of-the-art results by training the models on the publicly available irregular-shaped masks \cite{nv_irregular_maskdata} with models based on partial \cite{liu2018image} or gated convolutions \cite{yu2019free}.
In non-medical inpainting, the shapes are unpredictable, and training with irregular-shaped masks enables the inpainting models to reconstruct images of such nature; however, in the medical images, shapes and textures are somewhat predictable, and the irregular and rectangular masks do not correlate well with the texture and structure of the target, which is generally an organ or an organ-shaped area.

We hypothesize that the position and shape of the inpainting mask have an essential effect on image reconstruction performance. We verify this hypothesis through several experiments showing the impact of shape-aware mask generation and positioning. Positioning the masks inside the bounded region of an organ is also explored in \cite{tran21multi} and shown to improve the performance; nevertheless, they compare square masks to arbitrarily shaped masks in their work and show that the square masks achieve higher performance. The masks generated by our model are not arbitrary; instead, they are generated from pseudo-segments that conform to the distribution of the structures and shapes appearing in the image. Since our approach is dependent on the shape prior modelling for mask generation, rather than the pseudo-segment generation approach, despite the existence of few over-segmentation algorithms \cite{jampani2018superpixel,achanta2012slic,yang2020superpixel}, we use a fast and straightforward off-the-shelf method \cite{felzenszwalb2004efficient} that has been used in medical imaging \cite{ouyang2020self} for self-supervised segmentation\cite{farshad2022upsilon}. To evaluate the effect of different masking strategies, we rely on the image reconstruction metrics. We show that by conditioning the mask position and shape on pseudo-segments, the model's performance in image reconstruction improves significantly. 

To summarize our contributions: 1) We propose a self-supervised shape-aware mask localization and generation strategy for the image inpainting task that is more anatomy friendly, 2) We verify our hypothesis that the masks' location and shape are essential aspects of mask generation for image reconstruction in inpainting, 3) Our self-supervised mask generation method relies on randomly sampled pseudo-segments based on the superpixels in the image, 4) Our proposed method outperforms existing mask shapes (square and irregular) and random positioning of the mask on abdominal MR image reconstruction. 5) The source code for this work will be publicly released upon its acceptance.
\section{Related Works}
Image inpainting is known for refurbishing images in computer vision, but its applications extend to other tasks. In medical imaging, inpainting has unique applications, from anomaly detection to pretraining models for downstream tasks. Here we discuss some of the medical applications of inpainting.

Early works on medical image inpainting \cite{feng07var,mirko10automatic} focused on the removal of erroneous parts of the medical images, such as speckle noise or glossy appearance of the image. These works relied on classical methods such as Mumford-Shah function \cite{feng07var,mum89shah} or manual modification and smoothing of the pixels \cite{mirko10automatic}. 
Anomalies in organs can have a negative effect on the performance of the neural networks in other tasks such as image registration and segmentation; therefore, some works \cite{guizard15non,manjon2020blind} apply inpainting to remove these anomalies to increase the model robustness in other tasks. The effectiveness of inpainting in removing anomalies has been explored in \cite{guizard15non,manjon2020blind} for the removal of brain lesions and has been shown to improve brain atrophy detection performance \cite{guizard15non}. 

As a result of the advances in deep learning, recent inpainting methods focus on using deep neural networks for inpainting masked images \cite{tran20deep}. Armanious et al. propose an adversarial network for MR image inpainting \cite{armanious2019adversarial,aliakbari2015simulation}. In their next work \cite{armanious2020ipa}, they propose a new network for the inpainting of arbitrary regions. 

Due to the importance of edges and structure in the image, a couple of works \cite{feng07var,wang2021medical,tran21multi} aimed at enforcing the continuity of edges for inpainting. \cite{wang2021medical} uses edge and structure information in their inpainting framework for COVID CT, Abdominal CT, and Abdominal MR reconstruction to overcome the problem of distortion in medical images. In \cite{gapon2019medical}, multi-scale masks are used for medical inpainting by using large masks for homogeneous areas and small masks for structure details. Bukas et al. \cite{bukas2021patient} use inpainting for reconstruction and straightening of the broken vertebra to estimate the amount of cement needed for osteopathy. Another use-case of inpainting is the reconstruction of 3D brain MR scans from sparse 2D scans to save time \cite{kang2021deep}. Recently Kim et al. \cite{kim2021synthesis} proposed the generation of tumors in the healthy brain using inpainting to visualize the tumor spreading.
\section{Methodology}
\begin{figure}
    \centering
    \includegraphics[width=\textwidth]{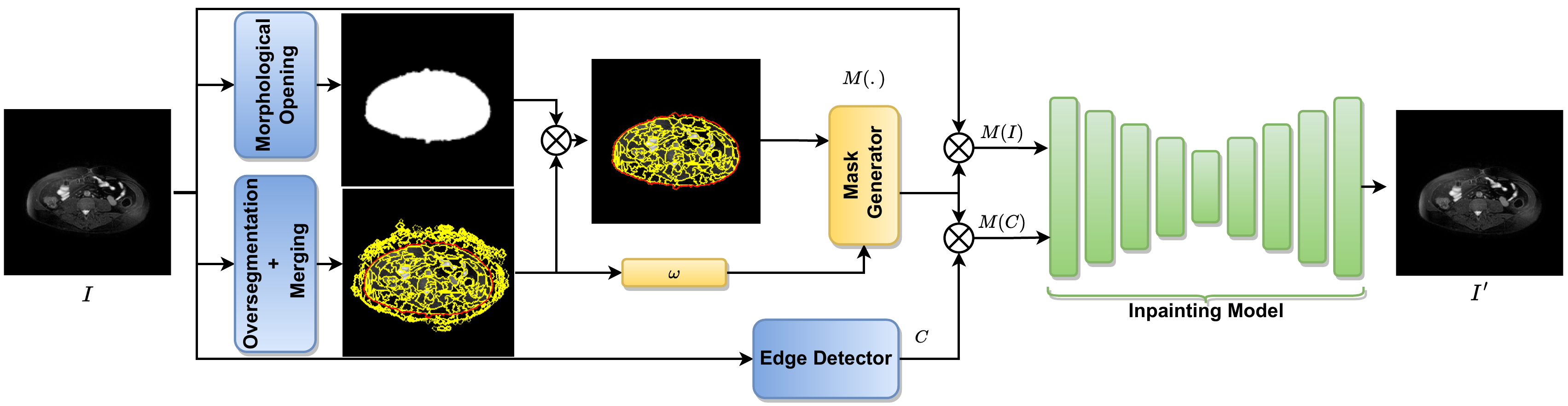}
    \caption{Here, we show an overview of our mask generation method. The organ bounded region and pseudo-segments are generated by morphological opening and the superpixel algorithm, respectively. Finally, our shape-aware masks are sampled using the mask generator and used for the inpainting model.}
    \label{fig:my_label}
\end{figure}

This section explains our proposed masking strategy for the inpainting task. Our methodology focuses on the mask configurations, which are the shape and the position of the masks used for training the inpainting model. We first define the inpainting framework and then present our proposed mask localization and mask shape generation techniques. We choose the recently published CTSDG \cite{guo2021image} model as our inpainting framework, which uses the edges in the image as an extra input and also has an independent objective function to ensure the continuity of the edges; hence, we show that the proposed masking strategy is the main contributing factor in the improvement of the model performance. An overview of our method is presented in \autoref{fig:my_label}.

\subsection{Inpainting framework}
Here, we define our inpainting framework, which is based on CTSDG \cite{guo2021image}. Since CTSDG relies on both the texture and the structure for image reconstruction, it also takes advantage of the edges in the image. The edges $C$ are obtained by applying a canny edge detector to the image. Given an input image $I \in \mathbb{R}$, the goal of our inpainting network $f(.)$ is to reconstruct pixels of the image which are masked by $M(.)$ before feeding it to $f(.)$. In addition to reconstructing the masked image, the inpainting model also reconstructs the masked edges $C'$. Therefore, the output of our network, which is the reconstructed image $I'$ becomes:
\begin{equation}
    I', C' = f(M(I), M(C))
\end{equation}

\subsection{Pseudo-segment Generation}
In order to generate pseudo-segments for our images, inspired by \cite{ouyang2020self}, we apply the Felzenszwalb superpixel over-segmentation algorithm \cite{felzenszwalb2004efficient} on all of the images in the dataset. The Felzenszwalb algorithm operates by mapping the image pixels into a graph defined by $G = (V,E)$, where $\nu \in V$ is the set of vertices and $\epsilon \in E$ are the edges in the graph connecting certain pairs of neighboring pixels. Each edge $\epsilon_{i,j}$ defines a connection between the neighbour pixels $\nu_i, \nu_j$ and a distance measure $\delta_{i,j} \in \Omega$ defines the dissimilarity between $\nu_i$ and $\nu_j$. Initially, the dissimilarity distances $\delta_{i,j}$ are computed for all the eight neighboring pixels of each pixel in the image. Then the dissimilarity measure $\Omega$ is used to merge the pixels for constructing the pseudo-segments. The sensitivity (size) of the generated pseudo-segments is defined by a predetermined threshold $\tau$ considering the distances between the vertices.

To avoid unwanted merging by the algorithm and to have more control over the superpixels' merging process, we set the minimum number of pixels in each superpixel to a relatively low value. Later based on the mean color and distance of the superpixels, they are merged to achieve a larger superpixel that we call pseudo-segments. The superpixel algorithm is designed such that each segment holds comparable information regardless of its number of pixels; we utilize this aspect to increase the likelihood of the appearance of such pseudo-segments in the final mask. We argue that since the non-perceptual objective functions for image reconstruction produce a scalar value that is usually based on the distance between the pixel values of $I$ and $I'$, the model's objective inherently becomes less sensitive to smaller segments that potentially hold information. Therefore, we intentionally increase the frequency of the pseudo-segments with a smaller number of pixels. To achieve that, we create a $2 \times N$ matrix $\omega$ called weight-biases. $N$ is set to the number of detected pseudo-segments in an image, and each pseudo-segment is paired with a value calculated by the inverse number of pixels in each pseudo-segment. This provides pseudo-segments $PS_n$ for each image $I_n$ used by the sampler to generate masks usable for the inpainting model.

\subsection{Masking Position}
We place a set of sampled pseudo-segments in specific regions to generate our desired mask. The placement regions are identified based on the centers of the pseudo-segments. Three approaches are outlined for the placement of the sampled pseudo-segments: 1) random locations in the image, 2) random locations inside the bounded region of the organ (IBR), 3) exact location of the pseudo-segment (OPS).

\subsubsection{Inside Bounded Region (IBR)} To obtain the bounded region of the organ, we apply morphological opening to the input image $I_n$. This provides us with a pseudo-segmentation mask that defines the whole region of the organ. The mask coordinates are then randomly sampled from the area inside the bounded region.

\subsubsection{On Pseudo-segments (OPS)} We hypothesize that the position of masked regions of the image for inpainting is an essential aspect for higher performance in image reconstruction. Therefore, we propose to use the center of randomly sampled pseudo-segments in each image as the position of our masked region. We argue that masking a whole region with a homogeneous structure would force the network to be more robust against overfitting and shortcut learning by using the information from the neighboring pixels. To find a set of positions to place our masks, we sample $m$ random pseudo-segments from each image and find the center coordinates of the pseudo-segments by mapping them into a bounding box and computing the center of the bounding box.

\subsection{Masking Shape}
In our framework, we propose using different shapes for the masks: 1) square, 2) irregular~\cite{nv_irregular_maskdata}, 3) PS: Our proposed shape-aware mask which relies on the pseudo-segments in the image.
Square shapes can have random sizes and locations, but to have comparable experiments, we reflect the size and locations of the pseudo-segments in the square masks. The irregular masks were obtained from the irregular shape dataset \cite{nv_irregular_maskdata,liu2018image}. We explain our proposed PS masks as follows.
\subsubsection{Pseudo-segment Masks}
Our pseudo-segments are generated by the Felzenszwalb \cite{felzenszwalb2004efficient} as mentioned in previous sections. In each training step, we randomly determine the value of $m$, and then with the likelihood of $\omega$, $m$ pseudo-segments are sampled. Then, by keeping the center of each mask fixed, we perform minimal random perturbations on each of the masks to enforce the model to learn the surrounding pixels of the mask.% With this mechanism, we have a vast collection of masks that also depend on the shape model of our domain.

\subsection{Objective Functions}
Our model is trained with a combination of reconstruction loss, perceptual loss, adversarial loss, and style loss similar to \cite{guo2021image}:

\begin{equation}
    \mathcal{L} = \| I' - I \| + \sum \| \phi (I') - \phi (I) \| + \sum \| \psi (I') - \psi (I) \| + \mathcal{L}_{adv} + \mathcal{L}_{int}
\end{equation}
where $\phi$ corresponds to the activation map of the image from a pre-trained VGG~\cite{simonyan2014very} network and $\psi$ is the Gram matrix of the activation map $\phi$ to preserve the style. The adversarial and intermediate losses are defined as $\mathcal{L}_{adv}$ and $\mathcal{L}_{int}$ respectively:
\begin{equation}
    \mathcal{L}_{adv} = \underset{f}{\min}\ \underset{D}{\max}\ \mathbb{E} [\log (D(I,C))] + \mathbb{E} [\log (1- D(I',C'))]
\end{equation}
\begin{equation}
    \mathcal{L}_{int} = BCE(C,C'') + \| I - I''\|
\end{equation}
where $C''$, $I''$ are intermediate reconstructions of the edge and image and $D(.)$ is the discriminator in our inpainting network.
\section{Experiments}
In this section, we present the experiments that validate our hypothesis. To evaluate our mask generation and positioning method, we train and test it on the recently published CHAOS \footnote{Link: https://chaos.grand-challenge.org/} \cite{CHAOS2021} (Combined CT-MR Healthy Abdominal Organ Segmentation) dataset. The CHAOS dataset includes healthy abdominal MR and CT scans from 80 patients. We train our model on T1-DUAL In-phase MR scans from 40 patients. The scans from $75\%$ of the patients were randomly chosen for the training and the scans from the remaining $25\%$ patients for evaluation.

\subsection{Experimental setup}
In our experiments, we follow the same protocols and hyperparameters as \cite{guo2021image} for the inpainting framework unless explicitly specified. All our models were trained for 10000 iterations, batch size of $6$, and a learning rate of $0.0002$ for the inpainting network and $0.1$ for the discriminator. All networks were trained with the Adam optimizer \cite{kingma2014adam}. For the superpixel over-segmentation, we used the scikit-image \cite{van2014scikit} library with the scale factor of 2, Gaussian filtering with $\sigma = 0.5$, and a minimum size of 9 for each segment. To merge the pseudo-segments, we set the threshold value $\tau = 10$ and refine the pseudo-segments based on their mean color and their distance to their neighbors. In all our experiments, we evaluate the model using three different masking techniques to show the model's robustness to masks of other shapes during testing. Since we did not observe any commonly used or publicly available benchmark for evaluating inpainting masks in prior works, we define our own evaluation protocol, including the square-shaped masks, which are the standard masks used in medical image inpainting.
Our three evaluation masks are: 1) segmentation shaped mask + OPS, which explicitly mask a segment that correlates with the shape of the organs, 2) multiple small squares, with similar total size to the area of the segments scattered in the image, and 3) one large random sized and randomly located square that masks $20\%$ to $60\%$ of the image to measure how the model can handle large consistent holes.
\subsubsection{Evaluation Metrics} We evaluate our method and the baselines on three different similarity metrics: 1) PSNR (Peak Signal-to-Noise Ratio), 2) SSIM \cite{wang2004image} (Structural Similarity Index Measure), and 3) LPIPS \cite{zhang2018unreasonable} (Learned Perceptual Image Patch Similarity). All the metrics were calculated only on the region of interest (RoI) where the masking had happened. 

\begin{table}[tb]
\caption{A comparison of various mask locations for training the model tested in 3 different evaluation settings. \textbf{IBR}: Inside Bounded Region, \textbf{OPS}: On Pseudo-Segments. The mask shape in these experiments is square.}
\resizebox{\textwidth}{!}{
\begin{tabular}{cccccccccc}
\hline
\multirow{3}{*}{Position} & \multicolumn{9}{c}{Evaluation Mask} \\ \cline{2-10}
& \multicolumn{3}{c}{Segments} & \multicolumn{3}{c}{Multiple Small Squares} & \multicolumn{3}{c}{Single Large Square} \\  \cline{2-10}
 & PSNR $\uparrow$ & SSIM $\uparrow$ & LPIPS $\downarrow$ & PSNR $\uparrow$ & SSIM $\uparrow$ & LPIPS $\downarrow$ & PSNR $\uparrow$ & SSIM $\uparrow$ & LPIPS $\downarrow$\\ \hline
Random & 16.66 & 0.60 & 0.2885 & 16.78 & 0.63 & 0.32 & \textbf{14.08} & 0.64 & 0.29 \\
IBR (Ours) & 18.04 & 0.67 & 0.2686 & 18.13 & 0.67 & \textbf{0.2586} & 13.65 & 0.65 & \textbf{0.2670} \\
OPS (Ours) & \textbf{18.07} & \textbf{0.68} & \textbf{0.2676} & \textbf{18.45} & \textbf{0.69} & 0.2888 & 13.49 & \textbf{0.67} & 0.2685 \\ \hline
\end{tabular} } \label{tab:pose}
\end{table}

\subsection{Results}
Our experiments are presented in three sections: \autoref{tab:pose} shows the effect of the masks locations on the performance of the model. In \autoref{tab:shape}, we discuss the shapes of the masks, and in \autoref{tab:last} we ablate the source and locations of the shape distributions and how our weight-biased sampling improves the results.

Finally, in \autoref{tab:last} we show the reconstruction performance of our proposed pseudo-segment masks on both IBR and OPS and masks that are sampled from another image to investigate the model performance's dependency on the domain of the mask shapes; however, since we do not use the same indices, the masks from the other image are sampled based on their own weight-biases.
\begin{table}[b]
\caption{A comparison of various mask shapes used for training the model tested in 3 different evaluation settings. The positioning of the masks for all shapes is based on OPS. \textbf{PS}: Pseudo-Segments}
\resizebox{\textwidth}{!}{
\begin{tabular}{cccccccccc}
\hline
\multirow{3}{*}{Shape} & \multicolumn{9}{c}{Evaluation Mask} \\ \cline{2-10}
& \multicolumn{3}{c}{Segments} & \multicolumn{3}{c}{Multiple Small Squares} & \multicolumn{3}{c}{Single Large Square} \\  \cline{2-10}
 & PSNR $\uparrow$ & SSIM $\uparrow$ & LPIPS $\downarrow$ & PSNR $\uparrow$ & SSIM $\uparrow$ & LPIPS $\downarrow$ & PSNR $\uparrow$ & SSIM $\uparrow$ & LPIPS $\downarrow$\\ \hline
Square & 18.07 & 0.68 & 0.2676 & 18.45 & 0.69 & 0.2888 & 13.49 & 0.67 & \textbf{0.2685} \\
Irregular & 16.44 & 0.61 & 0.3078 & 16.63 & 0.64 & 0.3174 & \textbf{14.13} & 0.66 & 0.2780 \\
PS (Ours)& \textbf{19.38} & \textbf{0.77}& \textbf{0.2655} & \textbf{19.58}& \textbf{0.78} & \textbf{0.2630} & 13.5 & \textbf{0.74} & 0.2853 \\ \hline
\end{tabular} } \label{tab:shape}
\end{table}

For the experiments in \autoref{tab:pose}, we first evaluate different mask localization approaches using the square masks to prove that regardless of the shape, appropriate mask positioning improves the reconstruction performance. It can be seen that bounding the masking region to the organ region brings the most significant improvement. Furthermore, if we enforce the mask centers to fall into the predicted pseudo-segments, the performance slightly improves. To ensure comparable results, we set the size of the squares the masks in all of the experiments to be the same.

%%%%%%%%%%%%%%%%%%%% 3 %%%%%%%%%%%%%%%%%%%%%%%%%%%%%
\begin{table}[htb]
\caption{\textbf{IBR}: Inside Bounded Region, \textbf{OPS}: On Pseudo-Segments, The shape of the masks in these experiments are based on pseudo-segments (PS). \textbf{Weighted}: refers to weighted sampling.}
\resizebox{\textwidth}{!}{
\begin{tabular}{cccccccccccc}
\hline
\multirow{3}{*}{Distribution}& \multirow{3}{*}{Weighted} & \multirow{3}{*}{Position} & \multicolumn{9}{c}{Evaluation Mask} \\ \cline{4-12}
& && \multicolumn{3}{c}{Segments} & \multicolumn{3}{c}{Multiple Small Squares} & \multicolumn{3}{c}{Single Large Square} \\  \cline{4-12}
 && & PSNR $\uparrow$ & SSIM $\uparrow$ & LPIPS $\downarrow$ & PSNR $\uparrow$ & SSIM $\uparrow$ & LPIPS $\downarrow$ & PSNR $\uparrow$ & SSIM $\uparrow$ & LPIPS $\downarrow$ \\
\hline
Same Image & - & OPS &18.25 &0.6456 &0.29 &18.5 & 0.6767&0.2955 & 14.02& 0.6865&0.2862 \\
Other Image & \Checkmark & OPS & 18.31 & 0.6295 & 0.3009 & 18.48 & 0.6552 & 0.3054 & \textbf{14.11} &0.6878 & 0.2871 \\
Same Image & \Checkmark & IBR & 19.16 & 0.6836 & \textbf{0.2544} & 19.18 & 0.7064 & \textbf{0.2599} & 13.42 &0.6619 & 0.29665 \\

Same Image & \Checkmark & OPS & \textbf{19.38} & \textbf{0.77} & 0.2655 & \textbf{19.58} & \textbf{0.78} & 0.263 & 13.5 & \textbf{0.74} & \textbf{0.2853} \\

\hline
\end{tabular} } \label{tab:last}
\end{table}

In \autoref{tab:shape}, we evaluate and compare our proposed pseudo-segment masks against the commonly used irregular \cite{nv_irregular_maskdata} and square masks. As it can be seen, the model trained with pseudo-segment shaped masks (PS) achieves the best performance in most of the evaluation settings and metrics.

It is shown in \autoref{tab:last} that the masks sampled from another image perform relatively better than the square shaped masks, which indicates the importance of the mask's shape; however, if the masks are sampled from the same image and positioned based on IBR, we achieve better performance. Finally, we observed in these experiments that the masks that used OPS and were sampled from the same image yield better overall performance.

\section{Discussion and Conclusion}
We proposed a self-supervised shape-aware mask generation method that relies on the pseudo-segments generated from the image. Our mask generation model considers both the shape and the location of the masks. Our proposed pseudo-segment shapes are more anatomy friendly than the commonly used square masks. We observed in our experiments that positioning the masks in the bounded region of the organ or on pseudo-segments has a significant effect on the model's reconstruction performance. Furthermore, we demonstrated that the model's reconstruction performance is heavily influenced by the shape of the masks used for training. Therefore, we conclude that the mask's position and shape should correlate with the shape of the parts in the image. Finally, we showed that using pseudo-segments from images other than the current image for mask generation has a higher performance than commonly used square shaped and irregular masks and therefore a dataset of our generated pseudo-segment masks can be considered a substitute to square and irregular masks for the medical image inpainting task.
\section*{Acknowledgement}
This work was supported in part by the Munich Center for Machine Learning (MCML) with funding from the Bundesministerium f\"ur Bildung und Forschung (BMBF) under the project 01IS18036B.

\bibliographystyle{splncs04}
\bibliography{main}

\end{document}